\documentclass[12pt]{article}
\title{No robust phases in aerogel:\\
$^3$He-A with orientational disorder in the Ginzburg-Landau model.
\\(Comment on papers by I.A. Fomin on robust phases)}
\author{G.E. Volovik\\
Low Temperature Laboratory,
Helsinki University of Technology\\
P.O.Box 2200, FIN-02015 HUT, Finland\\
and\\
L.D. Landau Institute for Theoretical Physics,
  Moscow\\
}

\begin{document}
\maketitle

\begin{abstract}
{In series of papers \cite{Fomin,FominResponse} Fomin introduced and
discussed the so-called robust phases in a system with
frozen orientational disorder (with application to superfluid $^3$He in
aerogel). We show that his consideration is based on the erroneous
overestimation of the fluctuation energy which comes from the interaction
of the Goldstone modes with the frozen disorder. This interaction leads
to the Imry-Ma effect, which destroys the orientational order, but is
unable to destroy the local structure of  $^3$He-A. There is no ground
for the robust phases.}
\end{abstract}

Following Fomin, let us discuss the superfluid $^3$He in aerogel using
the Ginzburg-Landau (GL) model supplemented by the interaction with the
frozen orientational disorder field $\eta_{ij}$:
\begin{equation}
F=F_0 +F_{grad} + F_{\eta}~.
\label{TotalGL}
\end{equation}
Here $F_0$ and $F_{grad}$ are condensation and gradient energies, and
\begin{equation}
F_{\eta}=\int \eta_{ij}({\bf r})A_{\mu i}({\bf r})
A^{*}_{\mu j}({\bf r})d^{3}r,
\label{1}
\end{equation}
where $\left<\eta_{ij}\right>=0$, and we only consider the orientational
anisotropy, i.e. the tensor $\eta_{ij}$ is traceless: $\eta_{ii}=0$. 

We assume that the disorder is relatively small. Therefore we can start
with  homogeneous states which have spatially uniform order parameter
$A_{\mu i}=A_{\mu i}^{(0)}=const$. Since  $\int \eta_{ij}d^{3}r=0$,
the energy of such state only comes from $F_0$:
\begin{equation}
F(A_{\mu i}^{(0)})=F_0(A_{\mu i}^{(0)})~.
\label{EnergyUniform}
\end{equation}
 We consider here the proper range of the parameters of the GL functional
$F_0$ (the $\beta$-parameters of 4-th order terms in $F_0$
\cite{VollhardtWoelfle}), for which  $^3$He-A has minimum energy. The
energy of the  uniform  $^3$He-A is smaller than the energy of any other
uniform phase by the magnitude
$\sim F_0\sim N_F\tau^2 T_c^2$, where $N_F$ is the density of states in 
 normal Fermi liquid, and $\tau
=1-T/T_c$.
The quasi-isotropic robust phase determined by the condition
$\eta_{ij}({\bf r})A^{(0)}_{\mu i} (A^{(0)}_{\mu j})^{*}=0$  \cite{Fomin},
has also higher energy. 

Let us consider now the second-order ($\eta^2$)
correction to the energy $F_0$. The  uniform  $^3$He-A is not the minimum
of the total GL functional (\ref{TotalGL}), that is why its
energy can be reduced by adding the non-uniform corrections
(fluctuations),
$A_{\mu i}=A_{\mu i}^{(0)}+a_{\mu i}$, with $a\propto \eta$ and thus 
$\left<a\right>=0$. 
The
$\eta^2$ terms contain the linear and quadratic terms in
$a_{\mu i}$. In ${\bf k}$ representation after diagonalization of the
$a^2$ terms one obtains
\begin{equation}
F=F_0(A_{\mu i}^{(0)}) + F_{\rm fl} ~,
\label{Symbolic}
\end{equation}
where the fluctuation energy:
\begin{equation}
F_{\rm fl}={1\over 2} \sum_{n, {\bf k}} a^2_{n, {\bf k}}\epsilon
_{n}({\bf k})  +  \sum_{n, {\bf k}}\tilde \eta_{n, {\bf k}}a_{n, {\bf k}} .
\label{GLCorrections}
\end{equation}
Here $\tilde \eta_{n, {\bf k}}$ comes from the product of $\eta$  and 
$A_{\mu i}^{(0)}$ matrices, and
$\epsilon _{n}({\bf k})$ is the spectrum of the $n$-th mode.  For
Goldstone modes (GM), 
\begin{equation}
\epsilon
_{G}({\bf k})\sim N_F\xi_0^2 k^2 ~ ,
\label{GolstoneSpectrum}
\end{equation}
and for other modes with gaps:
\begin{equation}
\epsilon
_{non-G}({\bf k})\sim N_F (\tau +k^2\xi_0^2)= N_F \tau(1
+k^2\xi^2) ~,
\label{NonGolstoneSpectrum}
\end{equation}
where $\xi=\xi_0/\sqrt{\tau}$ is the GL coherence length.

After minimization over $a$ one obtains
the contribution of fluctuations that reduce
the $^3$He-A energy:
\begin{equation}
F_{\rm fl}=-{1\over 2} \sum_{n, {\bf k}} \tilde
\eta^2_{n, {\bf k}}
\epsilon^{-1} _{n}({\bf k}) ,
\label{EnergyCorrections}
\end{equation}
There is no
divergence at small $k$, and the integral is concentrated at large $k
\gg 1/\xi$, if we assume that the frozen disorder
is concentrated at $1/\xi_0> k\gg 1/\xi$. It mainly gives the
shift of the transition temperature $T_c$. Actually the traceless
orientational disorder increases the transition temperature.  Subtracting
from Eq.(\ref{EnergyCorrections}) the integral 
with
$\tau=0$ in the denominator, one obtains the integral
$\propto \int d^3k /k^2(1+k^2\xi^2)$ concentrated at
$k\sim\xi^{-1}$:
\begin{equation}
\Delta F_{\rm fl}  \sim 
(A^{(0)})^2 \frac {\eta^2_{0} }{\tau \xi^3 N^2_F}\sim \alpha F_0~,
\label{Subtracted}
\end{equation}
where $\eta^2_{0}=\int d^3r \left<\eta({\bf r}) \eta(0)\right>$ and
$\alpha$ is the Larkin-Ovchinnikov parameter \cite{LO}
\begin{equation}
 \alpha = \frac {\eta^2_{0} }{\tau^{1/2} \xi_0^3 N^2_F}\ll 1~.
\label{alpha}
\end{equation}

We can already stop at this point, since the fluctuation energy is
small compared to the condensation energy, and thus $^3$He-A remains the
only possible phase. However, Fomin points out that the interaction of the
frozen disorder with GM changes the situation, because due to these modes
the amplitude of fluctuations of the non-robust states diverges at small
$k$:
$\left<a^2\right>\propto \int d^3k/k^4 \sim \int dk/k^2 \sim L$, where
$L^{-1}$ is the infra-red cut-off parameter. This gives
\begin{equation}
\left<a^2\right>    \sim  \frac{ \alpha L}{\xi} (A^{(0)})^2~   .
\label{AmplD}
\end{equation}
At $L\sim \xi$, fluctuations are small if $\alpha\ll 1$, and this is
the condition for the applicability of the GL approach. But fluctuations 
become comparable to $A^{(0)}$ at
\begin{equation}
L \sim  \frac{\xi}{\alpha}\gg \xi~,
\label{ImryMaLength}
\end{equation}
and this scale $L$ provides the infrared cut-off.

This consideration is certainly true, but it is the well known Imry-Ma
effect
\cite{ImryMa}:  Since the
Eq.(\ref{AmplD}) describes the fluctuations of the GM, it corresponds to
the change in the orientation of the order parameter $A$ without
disturbing its structure. The scale
$L$ at which
$\left<a^2\right>  \sim  (A^{(0)})^2$ thus indicates the scale at which
the orientation of $A$ changes by angle of order
$\pi/2$.  This is just the Imry-Ma length scale. The state
looses the orientational long-range order due to interaction of the GM with
the frozen orientational disorder. The similar destruction of the
long-range translational order in the mixed state of superconductors by
inhomogeneities was found even earlier
\cite{Larkin}. The Imry-Ma effect applied to 
$^3$He-A in aerogel was discussed in \cite{GlassState}.

Fomin claims that the GM also leads to the divergent contribution
to fluctuation energy, which is absent in the robust phases.
Let us see. The contribution of the  GM with wavelength $L$
to the fluctuation energy
$F_{\rm fl}$ in Eq. (\ref{EnergyCorrections}) is proportional to
$\int_0^{1/L} k^2dk/k^2 \sim 1/L$. The fluctuation energy in
Eq.(\ref{Subtracted}) comes from scale $\xi$ and is proportional to
$\int  k^2dk/(k^2 +1/\xi^2)\sim
\int_0^{1/\xi} dk
\sim 1/\xi$. Thus the contribution of GM with wavelength $L$ is by
factor
$\xi/L=\alpha$ smaller, and gives
the second-order in
$\alpha$ correction to the GL energy. This is just the Imry-Ma energy
gain due to the orientational disorder of the order parameter:
\begin{equation}
F_{\rm Imry-Ma} \sim \alpha \Delta F_{\rm fl} 
\sim 
\alpha^2 F_0 \ll F_0~.
\label{ImryMa}
\end{equation}
At the  Imry-Ma wavelength $L$, the interaction with the frozen disorder
is on the order of the gradient energy \cite{ImryMa}. Thus the
contribution of GM with  wavelength
$L\gg \xi$ to the energy is on the order of the gradient energy
at this scale, and thus contains  the small factor $\xi^2/L^2$ compared to
the condensation energy $F_0$. This is demonstrated in Eq. (\ref{ImryMa}),
since
$\xi/L=\alpha \ll 1$.

The equation (\ref{ImryMa}) contradicts to the statement by Fomin
\cite{FominResponse}, who erroneously concludes that the contribution of
GM contains the large factor
$1/\alpha$   compared to the contribution of the non-Goldstone
modes:
$F_{fl-G}\sim \alpha^{-1}F_{fl-non-G}$, and thus, due to
GM, the fluctuation  energy
is comparable to the condensation energy: $F_{fl-G}\sim F_0$. This
provides the justification for introduction of the robust phases where the
disorder does not interact with GM, and thus there is no 
divergence in the amplitude of the order parameter. This justification is
wrong and thus there is no basis for the robust phase.
 The same conclusion was made by Mineev and Zhitomirsky in their Comment
 \cite{MineevZhitomirsky}.

In conclusion, the Goldstone modes, i.e. fluctuations in the
direction of the degeneracy of the order parameter, do lead to the
divergence of the amplitude of the order parameter. But their
contribution to energy does not experience any divergence and is small
compared to the condensation energy by the parameter $\alpha^2\ll 1$. This
is nothing but the Imry-Ma effect, which leads to disorder in the
orientation of the order parameter at large length $L=\xi/\alpha\gg \xi$
without changing the local structure of the order parameter. Since the
condensation energy
$F_0$ is dominating, the local order parameter must be in the  $^3$He-A
state everywhere (at least within the GL model (\ref{TotalGL})). The
robust phase is not the extremum of $F_0$, and thus is not the solution
of GL equations. Thus within the GL model
with the frozen orientational disorder the Imry-Ma approach
is valid and it does not leave any room for the robust phase.

I thank I.A. Fomin, N.B. Kopnin and V.P. Mineev for discussions.

\end{document}